\documentclass[reprint, secnumarabic,amssymb, nobibnotes, superscriptaddress, aps, prl]{revtex4-1}
\usepackage{graphicx}
\usepackage[colorlinks=true,citecolor=blue,linkcolor=blue,urlcolor=blue]{hyperref}

\begin{document}

\title{Observation of a Charge Density Wave Incommensuration Near the Superconducting Dome in Cu$_{x}$TiSe$_2$}%

\author{A. Kogar}
\email{kogar2@illinois.edu}
\affiliation{\mbox{Department of Physics and Materials Research Laboratory, University of Illinois, Urbana, IL, 61801, USA}}
\affiliation{\mbox{Materials Science Division, Argonne National Laboratory, Argonne, Illinois, 60439, USA}}
\author{G.A. de la Pena}
\author{Sangjun Lee}
\author{Y. Fang}
\author{S.X.-L. Sun}
\affiliation{\mbox{Department of Physics and Seitz Materials Research Laboratory, University of Illinois, Urbana, IL, 61801, USA}}
\author{D. B. Lioi}
\author{G. Karapetrov}
\affiliation{\mbox{Department of Physics, Drexel University, Philadelphia, Pennsylvania, 19104, USA}}
\author{K.D. Finkelstein}
\author{J.P.C. Ruff}
\affiliation{\mbox{Cornell High Energy Synchrotron Source, Cornell University, Ithaca, New York, 14853, USA}}
\author{P. Abbamonte}
\affiliation{\mbox{Department of Physics and Seitz Materials Research Laboratory, University of Illinois, Urbana, IL, 61801, USA}}
\author{S. Rosenkranz}
\affiliation{\mbox{Materials Science Division, Argonne National Laboratory, Argonne, Illinois, 60439, USA}}
\date{\today}

\begin{abstract}
X-ray diffraction was employed to study the evolution of the charge density wave (CDW) in Cu$_x$TiSe$_2$ as a function of copper intercalation in order to clarify the relationship between the CDW and superconductivity. The results show a CDW incommensuration arising at an intercalation value coincident with the onset of superconductivity at around $x$=0.055(5). Additionally, it was found that the charge density wave persists to higher intercalant concentrations than previously assumed, demonstrating that the CDW does not terminate inside the superconducting dome. A charge density wave peak was observed in samples up to $x$=0.091(6), the highest copper concentration examined in this study. The phase diagram established in this work suggests that charge density wave incommensuration may play a role in the formation of the superconducting state.
\end{abstract}

\maketitle

The discovery of superconductivity in 1T-TiSe$_2$ upon copper intercalation led to a renaissance in the study of superconductivity in the transition metal dichalcogenides (TMDs)\cite{morosan}. Numerous studies followed this initial work partly because the phase diagram presented was reminiscent of other unconventional superconductors. With increasing copper intercalation, the charge density wave in Cu$_x$TiSe$_2$ is suppressed and a superconducting dome appears, centered around $x=0.08$. A follow-up study of 1T-TiSe$_2$ under pressure exhibited a similar phase diagram, lending credence to the idea of a quantum critical point inside the superconducting dome\cite{kusmartseva}.

Since these initial investigations, however, a more complex picture has arisen. Subsequent studies of 1T-TiSe$_2$ and other TMDs have suggested a more subtle relationship between the charge density wave (CDW) and superconducting orders, with indication that incommensuration and disorder effects may be important\cite{liNature, joe, Liu, sipos, petrovic}. In particular, it has been observed that many TMD compounds exhibit a coexistence of superconductivity with incommensurate charge density wave order, but the coexistence of superconductivity and commensurate CDWs are relatively rare.

Prominent members of the former class include, but are not limited to, 2H-NbSe$_2$, 1T-TaS$_{2-x}$Se$_x$, 2H-TaS$_2$, as well as pressure-tuned 1T-TaS$_2$, where the maximum superconducting transition temperatures are, respectively, 7.2K, 3.6K, 0.8K, and 5K\cite{Gabovich, Liu, Wagner, sipos}. A well-known member of the latter group is 2H-TaSe$_2$, in which a 3x3x1 commensurate CDW and superconductivity coexist at a comparatively low 133mK\cite{yokota}. 

Notably, in 1T-TaS$_2$, the commensurate CDW is destabilized with pressure or lithium ion intercalation, but an incommensurate CDW survives and coexists with superconductivity\cite{sipos, feng2015}. Likewise, in 1T-TaS$_{2-x}$Se$_x$, superconductivity only appears between two regions of commensurate CDWs as $x$ is varied, again coexisting only with the incommensurate CDW\cite{Liu}. In the case of 2H-TaSe$_2$, the superconducting transition temperature can be raised to ~2K by irradiating the sample, in effect disrupting the CDW commensuration and introducing disorder\cite{mutka}.

In this work, we revisit the prototypical Cu$_x$TiSe$_2$ system to more precisely determine the charge density wave phase diagram based on X-ray diffraction. Cu$_x$TiSe$_2$ has previously been shown to host superconductivity from $x$=0.04$-$0.10\cite{morosan}. Pure 1T-TiSe$_2$ is notable among the transition metal dichalcogenides, because it is the only known TMD that undergoes a transition directly to a commensurate CDW state without exhibiting an intermediate incommensurate phase. In a recent high-pressure X-ray diffraction study, it was shown that CDW domain walls develop above the superconducting dome in this compound\cite{joe}. These domain walls led to the observation of a slight incommensuration in the CDW wavevector away from the 2$a\times$2$a\times$2$c$ superstructure. Furthermore, the CDW persisted to much larger pressures than suggested by the study based on transport, invalidating the idea of a quantum critical point inside the superconducting dome \cite{joe, kusmartseva}. Because copper-intercalated TiSe$_2$ also exhibits a superconducting dome, it is natural to ask whether such an incommensuration develops with the addition of copper intercalants and if the CDW persists to high intercalation values as well.

In order to investigate these possibilities, X-ray diffraction experiments were undertaken at beamlines C1 and A2 at the Cornell High Energy Synchrotron Source (CHESS), as well as on a lab-based X-ray source. Experiments at C1 beamline at CHESS were performed with the use of an energy-resolving silicon drift detector with an energy resolution of $\sim$1keV. Through the use of a Si(220) monochromator, some of the first harmonic signal was filtered into the incident beam, so that the beam possessed both 12keV and 24keV photons. A single scan therefore measured the (1/2,1/2,7/2) and (1,1,7) peaks simultaneously with the 12keV and 24keV photons respectively\cite{joe}. This technique made possible a study of the CDW commensuration to a high degree of accuracy, inherently correcting for sample misalignments. Because the energy-discriminating capability of the detector is coarse, it is still the equal-time density correlation function, $S(\textbf{q})$, that is being measured to a good approximation in the experiment. 

Cu$_x$TiSe$_2$ samples were grown using iodine vapor transport methods described in Refs.\cite{karapetrov, growth}. The samples had intercalation values ranging from $x=0-0.091(6)$. The elemental composition of the samples was determined by energy-dispersive X-ray spectroscopy (20x20 $\mu$m$^2$ beam spot) on multiple regions of each sample to ensure homogeneity. Only five of the nine samples were measured at the C1 beamline, where the harmonic filtering technique was possible. Hence, commensuration measurements were only conducted on these samples. For all measurements, Cu$_x$TiSe$_2$ crystals were mounted on an aluminum sample holder and cooled with a closed-cycle cryostat to a base temperature of 11K.

\begin{figure}
	\includegraphics[scale=0.31]{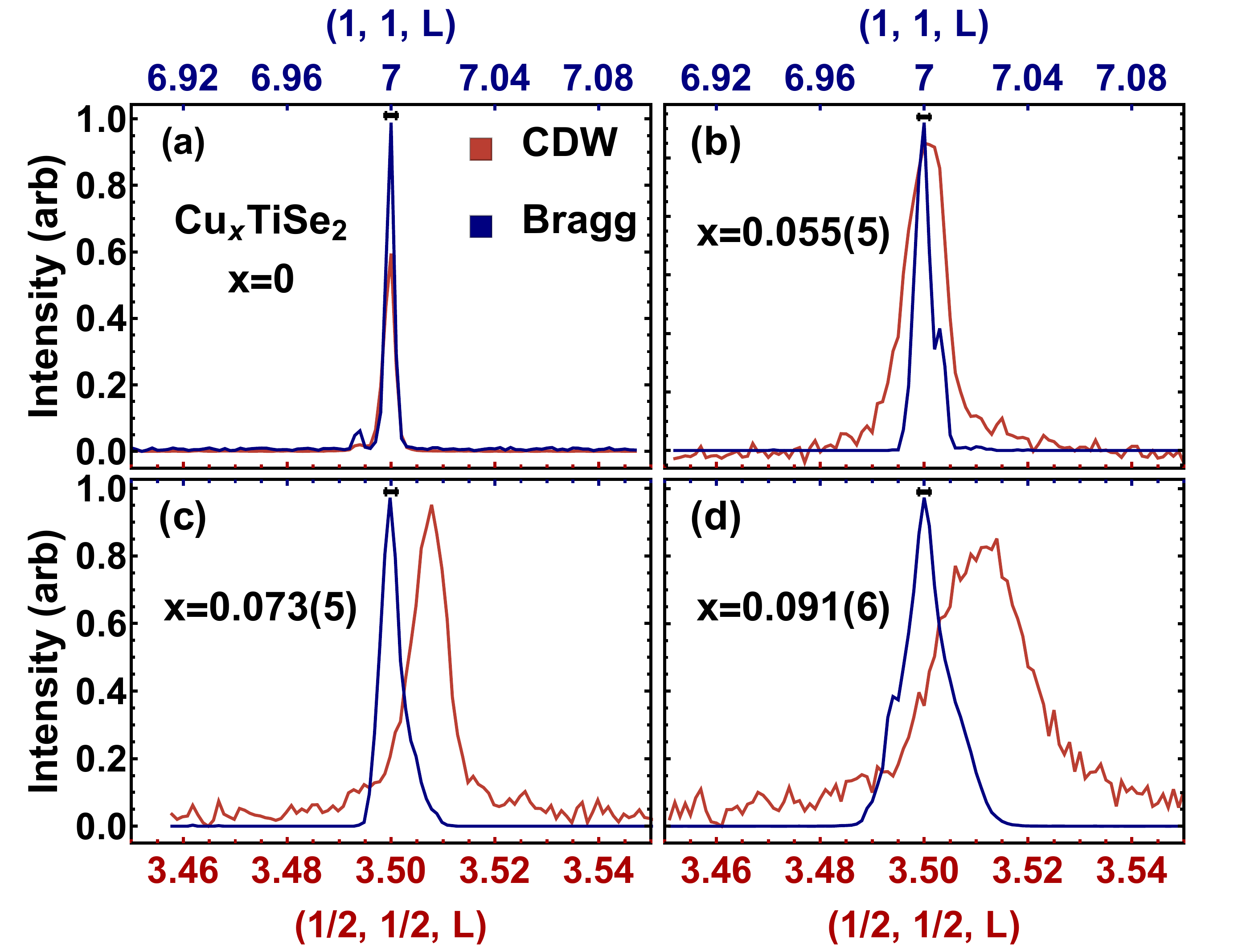}
	\caption{(color online) (a)-(d) Simultaneous L-cuts of the (1,1,7) Bragg peak as well as the (1/2,1/2,7/2) charge density wave superlattice peak for various amounts of copper intercalation in Cu$_x$TiSe$_2$. Data in panels (a)-(d) were taken at 50K, 11K, 15K and 20K respectively. Momentum resolution is indicated with the black line above the peaks.}
\label{fig:incommDoping}
\end{figure}

Fig.\ref{fig:incommDoping}(a)-(d) shows representative cuts along the L-direction for four different intercalant concentrations. The plots illustrate the capability of the harmonic filtering technique to clearly identify small deviations away from perfect commensuration by comparing the locations of the Bragg and charge density wave superlattice peaks in reciprocal space. Fig.\ref{fig:incommDoping}(a) shows resolution-limited peaks in L-scans for both the (1,1,7) and (1/2,1/2,7/2) peaks in the pure TiSe$_2$ sample at 50K. Within our experimental resolution, the charge density wave is perfectly commensurate with the lattice. However, for samples with $x\geq$0.055(5) copper intercalation, a clear incommensuration, $\delta$, is observed in the charge density wave along the L-direction [Figs.\ref{fig:incommDoping}(c) and (d)]. The incommensuration was observed to be temperature-independent within experimental error.

\begin{figure}
	\includegraphics[scale=0.42]{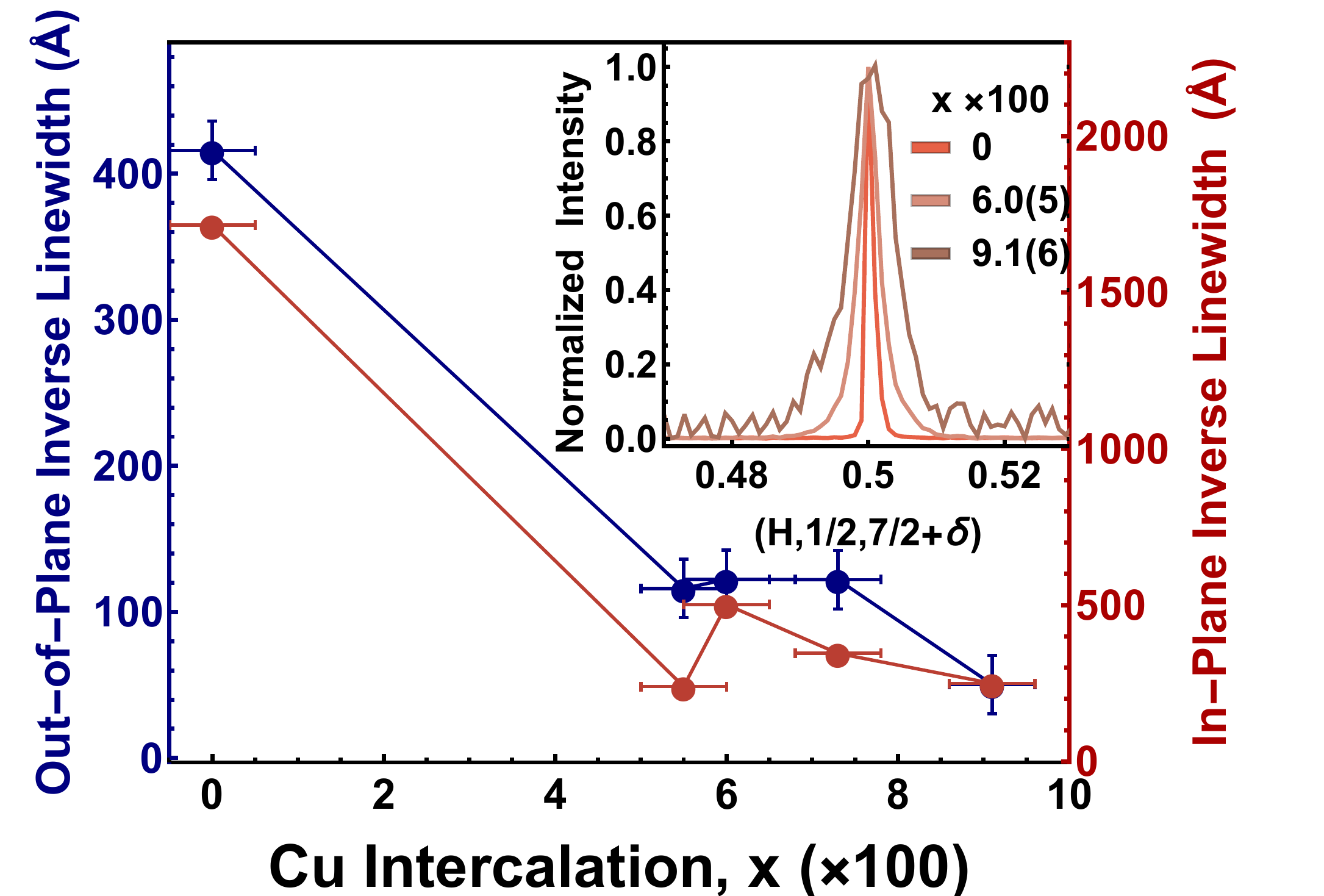}
	\caption{(color online) In-plane and out-of-plane inverse linewidths as a function of copper intercalation, obtained from Gaussian fits to the H- and L-cuts of the (1/2,1/2,7/2+$\delta$) superlattice peak. (Inset) H-cuts of the Cu$_x$TiSe$_2$ CDW superlattice peak at 50K, 13K and 20K for $x$=0, 0.060(5) and 0.91(6) respectively.}
\label{fig:corrLength}
\end{figure}

\begin{figure}
	\includegraphics[scale=0.4]{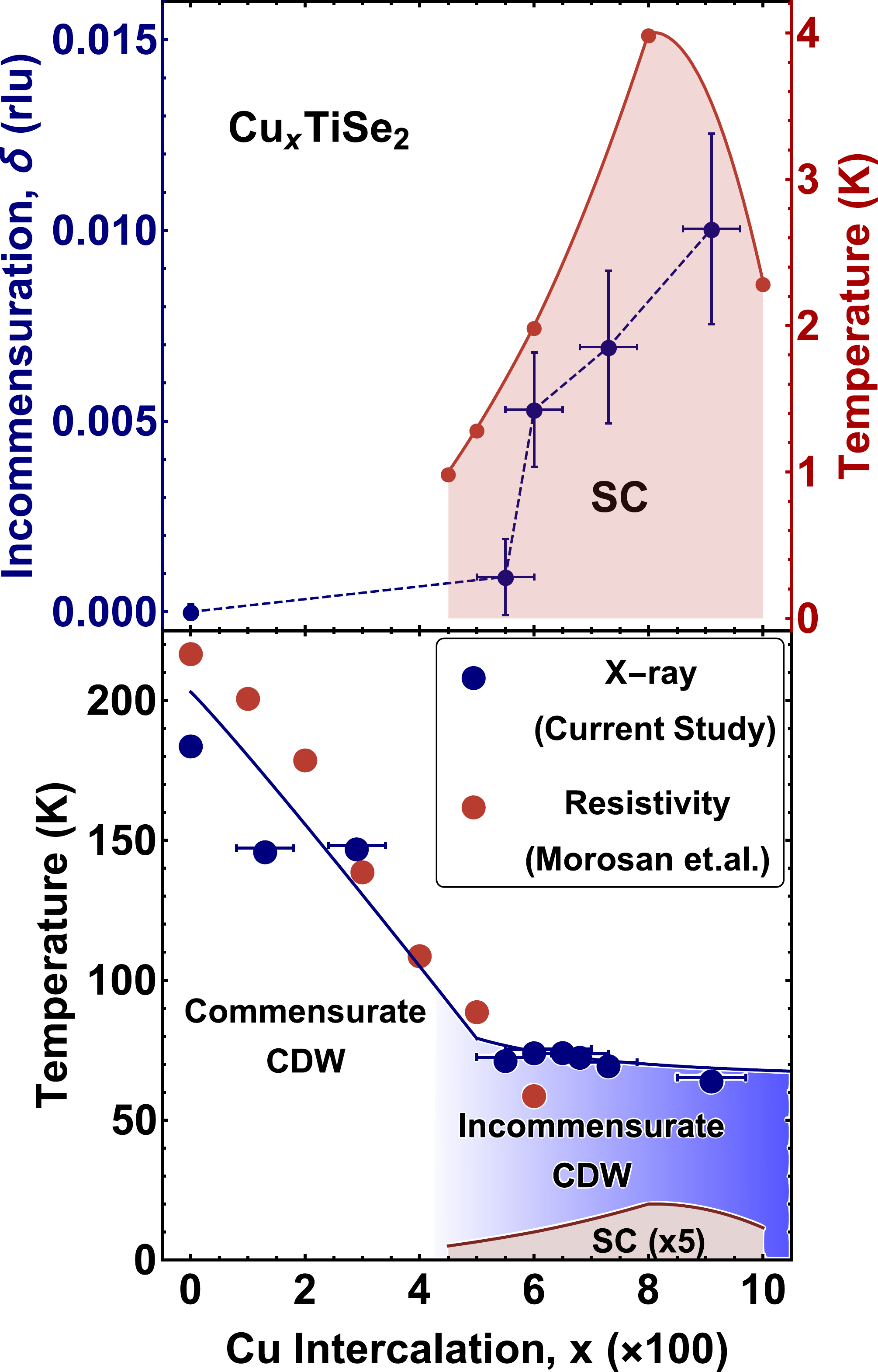}
	\caption{(color online) (Top) The degree of incommensuration along L as a function of copper intercalation. The red points outlining the superconducting dome are taken from Ref.\cite{morosan}. (Bottom) The phase diagram for Cu$_x$TiSe$_2$, where the blue points were obtained in this study. The red points and the line delineating the superconducting region are from Ref.\cite{morosan}.}
\label{fig:phase}
\end{figure}

Interestingly, the CDW incommensuration only becomes distinguishable around the same intercalation content at which superconductivity emerges, as shown in the top panel of Fig.\ref{fig:phase}. This is similar to the behavior observed for TiSe$_2$ under pressure, where incommensurate fluctuations were observed above the superconducting dome\cite{joe}. We attribute the existence of the CDW incommensuration to the presence of domain walls or stacking faults in the charge density wave along the c-axis. From studies of the nearly commensurate phase of 1T-TaS$_2$\cite{wuLieber}, it is known that ordered domain walls can give rise to a slight incommensuration in the CDW wavevector. We estimate that the stacking faults here occur at an average of $\sim$1/(2$\pi \times\delta$) = 1/(2$\pi \times$0.007) $\approx$ 22 lattice units along the c-axis for the $x$=0.073(6) sample, which is near the optimal intercalation for superconductivity.

It is worth comparing this value to the out-of-plane Ginzburg-Landau (GL) coherence length within the framework for anisotropic superconductors\cite{tinkham}. The in-plane and out-of-plane GL coherence lengths for Cu$_{0.07}$TiSe$_2$ and Cu$_{0.085}$TiSe$_2$ have been obtained from measurements of the upper critical fields, giving $\xi_{||}\sim$213$\textrm{\AA}$ and $\xi_{\perp}\sim$125$\textrm{\AA}$\cite{morosanPRB, growth}. This translates to $\xi_{||}\sim$60 in-plane lattice units and $\xi_{\perp}\sim$21 out-of-plane lattice units respectively. Although only an order of magnitude estimate, the agreement between the out-of-plane coherence length and stacking fault period is marked and suggestive of a relationship between the incommensuration of the CDW and superconductivity. In particular, it indicates that there may be an optimum stacking fault period for superconductivity.

\begin{figure}
	\includegraphics[scale=0.4]{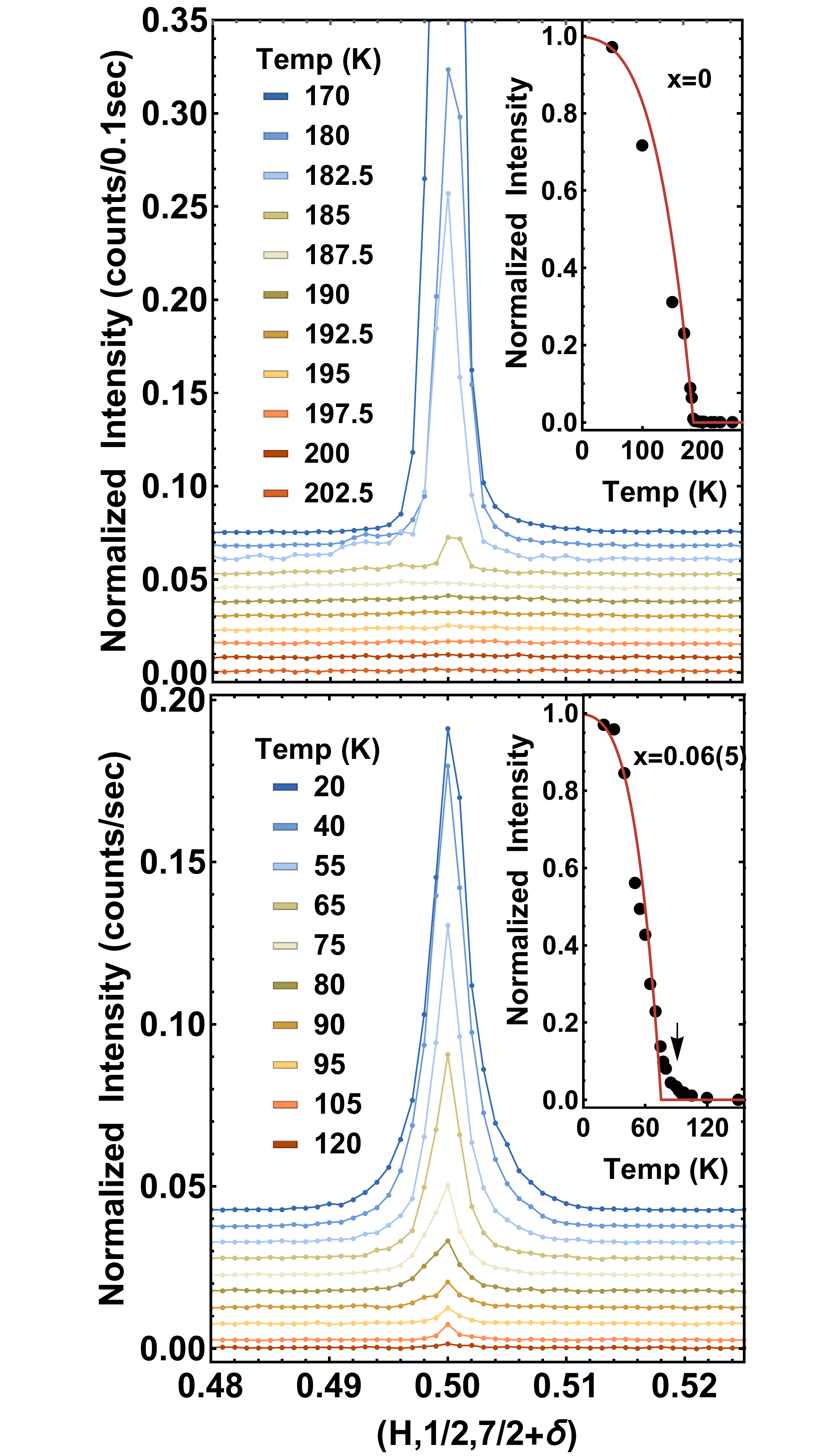}
	\caption{(color online) H-cuts of the CDW superlattice peak at several temperatures for pure TiSe$_2$ (top) and of Cu$_x$TiSe$_2$ with $x=$0.060(5) (bottom). Intensities are normalized to the incident beam counts. (Insets) Black dots indicate the integrated intensity of the H-cuts at various temperatures. The red line is the fit used to obtain the transition temperature.}
\label{fig:tempDep}
\end{figure}  

If we follow this line of reasoning further, the in-plane GL coherence length implies an in-plane incommensuration of $\delta_{||}\sim 0.002$. When taking into account peak linewidths and instrumental resolution, the predicted in-plane incommensuration would barely be observable within experimental error. Indeed, such an in-plane incommensuration was not observed in this study within these limitations [Fig.\ref{fig:corrLength} inset]. Nonetheless, it should be noted that the existence of in-plane domain walls has been inferred in a recent study using a gate-tuned thin film of pure 1T-TiSe$_2$\cite{liNature}. In that study, the domain walls were also thought to play a crucial role in the development of superconductivity.

The data in Fig.\ref{fig:incommDoping} is also notable in that the charge density wave is shown to survive despite significant copper intercalation. Previous studies based on transport measurements were not able to determine whether the charge density wave persisted above intercalation values of $x\approx$0.06\citep{morosan}. True long-range order does not develop in the intercalated compounds, however, as seen in Fig.\ref{fig:incommDoping}(b)-(d). Apart from the pure TiSe$_2$ sample, the CDW peaks observed were not resolution-limited, and the linewidths were broad for the highly intercalated samples, particularly along L. It is important to note also that the linewidth of the main Bragg peaks also broaden considerably along L, indicating considerable crystallographic disorder with increasing copper intercalation. The peaks were fit with a Gaussian profile along H and L to obtain the in-plane and out-of-plane inverse linewidths.

The evolution of the inverse linewidths of the CDW peak with intercalation content is shown in Fig.\ref{fig:corrLength}. To obtain correlation lengths from linewidths requires a detailed deconvolution of the instrumental resolution, and was not undertaken in this work. However, the inverse linewidths can serve as an estimate of the correlation length, especially along L, where the linewidths for the intercalated samples far exceeded the instrumental resolution. Along H, the inverse linewidths can still be used as an estimate for the in-plane correlation length, but because the linewidths are comparatively narrow [Fig.\ref{fig:corrLength} inset], they only strictly provide a lower bound. The in-plane and out-of-plane inverse linewidths for the $x=0.91(6)$ compound were approximately 250$\textrm{\AA}$ and 50$\textrm{\AA}$ respectively, yielding a correlation of $\sim$70 in-plane lattice units and $\sim$8 out-of-plane lattice units. For comparison, the pure TiSe$_2$ samples exhibited an in-plane and out-of-plane correlation of at least 485 and 70 in-plane and out-of-plane lattice units respectively. These observations indicate that although the resistivity anomaly is no longer visible for the highly intercalated samples\cite{morosan}, the CDW changes only quantitatively with greater intercalation. The CDW persists even in the sample with the highest intercalation content examined in this study.

In order to acquire a full phase diagram outlining the nature of the CDW, it was necessary to measure the temperature dependence of the CDW superlattice peaks for each sample. Fig.\ref{fig:tempDep} shows H-cuts of the (1/2,1/2,7/2+$\delta$) peak at a selection of temperatures for two representative samples, the pure and $x=$0.060(5) compound. The intensities in Fig.\ref{fig:tempDep} are normalized to the incident counts measured using an ion chamber situated upstream from the sample. To obtain the CDW transition temperatures, T$_{CDW}$, the integrated intensity of the H-cut was plotted against the temperature and fit with the universal fit function defined in the supplement of Ref.\citep{joe}. The fits were found to be excellent, as shown in the insets of Fig.\ref{fig:tempDep}. The transition temperature of the pure sample was found to be 185K, which is lower than most quoted transition temperatures in the literature\cite{diSalvo}. We attribute this to the presence of selenium vacancies, which has been shown to decrease the observed transition temperature\cite{diSalvo}. While the transition into the CDW state was sharp for the pure TiSe$_2$ sample [Fig.\ref{fig:tempDep} top panel inset], all intercalated samples exhibited a rounding of the transition due to the presence of disorder. The rounding is indicated with an arrow in the bottom panel inset of Fig.\ref{fig:tempDep}. In these samples, the fit function was therefore used to define the transition temperature, as shown with the red curve, despite the presence of a residual CDW peak above T$_{CDW}$.

The main results of this paper are summarized in Fig.\ref{fig:phase}. The bottom panel of Fig.\ref{fig:phase} shows the phase diagram obtained from these studies, including both the temperature dependence and incommensuration of the CDW. Data from Ref.\cite{morosan} is also included for comparison. It is immediately recognizable that there exist two regions of the phase diagram, with T$_{CDW}$ decreasing markedly from $x=0-0.055(5)$ and then leveling out for higher intercalation concentrations. This is in contrast to the phase diagram obtained in pressure-tuned studies of TiSe$_2$, where the observed CDW phase boundary decreases monotonically with increasing pressure\cite{joe, kusmartseva}.

We suggest that the segregation of the phase diagram occurs because copper intercalants electron-dope the Ti-3$d$ conduction band\cite{qian}, leading to a suppression of excitonic correlations. Many studies have advanced that both electron-phonon coupling and electron-hole coupling play significant roles in driving the charge density wave transition in pure 1T-TiSe$_2$\cite{diSalvo,kidd,cercellier,hellmann,rohwer, rossnagel,weber, porer,vanWezel, vanWezelprb}. Electron-doping selectively weakens the excitonic contribution to the charge density wave by shifting the chemical potential into the conduction band, thus enhancing screening effects, while leaving the electron-phonon interaction less affected. This interpretation provides a natural explanation to the phase diagram phenomenology: both excitonic and electron-phonon interactions contribute to driving the CDW transition in the low-intercalation region, while only electron-phonon coupling is relevant in the high-intercalation region. This picture enables us to approximate that pure 1T-TiSe$_2$ receives at most a T$_{CDW}$($x=0$)$-$T$_{CDW}$($x \approx 0.091(6))$ $\approx$185K-65K$\approx$120K boost in T$_{CDW}$ due to the presence of excitonic correlations.

To conclude, we have observed an incommensuration of the CDW in Cu$_x$TiSe$_2$ along the L direction arising at an intercalant concentration which coincides with the onset of superconductivity. This result appears to corroborate an increasing number of experiments demonstrating the importance of CDW incommensuration in the development of superconductivity in the TMDs\cite{joe,Liu,sipos,mutka}. Crystallographic disorder, however, may also be contributing to the development of superconductivity, a scenario we cannot rule out in Cu$_x$TiSe$_2$. In addition, we have shown that the CDW survives up to larger intercalant concentrations than previously thought, rather than terminating near or inside the superconducting dome.

We gratefully acknowledge J. van Wezel for fruitful discussions. Synchrotron experiments were supported by the Materials Sciences and  Engineering Division, Basic Energy Sciences, Office of Science, U.S. DOE. Lab-based x-ray experiments were supported by DOE-BES grant no. 
DE-FG02-06ER46285. Growth of TiSe$_2$Cu$_x$ crystals was supported by 
NSF grant ECCS-1408151. CHESS is supported by the National Science 
Foundation and the National Institutes of Health/National Institute of 
General Medical Sciences under NSF award DMR-1332208. P.A. acknowledges support from the Gordon and Betty Moore Foundation's EPiQS initiative through grant GBMF4542.

\bibliography{bib}
\end{document}